\documentclass[%
 reprint,
 amsmath,amssymb,
 aps,
]{revtex4-2}
\usepackage{booktabs}
\usepackage{enumitem} 
\usepackage{gensymb} 
\usepackage{upgreek}

\usepackage{hyperref}
\usepackage{lineno}
\usepackage{mathrsfs}

\usepackage{graphicx}
\usepackage{dcolumn}
\usepackage{bm}

\begin{document}

\preprint{APS/123-QED}

\title{A quantitative performance analysis of two different interferometric\\ alignment sensing schemes for gravitational wave detectors}


\author{Raed Diab}
 \altaffiliation{contact@raeddiab.com}
\author{Alvaro Herrera}
\author{Paul Fulda}%
\affiliation{University of Florida, Gainesville, Florida 32611, USA}%


\author{Chance Jackson}
\affiliation{Syracuse University, Syracuse, New York 13210, USA}


\date{\today}

\begin{abstract}

Precise laser alignment in optical cavities is essential for 
high-precision laser interferometry. We report on a table-top optical 
experiment featuring two alignment sensing schemes: the conventional 
Wavefront Sensing (WFS) scheme which uses quadrant photodetectors (QPDs) to recover optical alignment, and the newly developed Radio Frequency 
Jitter Alignment Sensing (RFJAS) scheme, which uses an electro-optic beam 
deflector (EOBD) to apply fast angular modulation. This work evaluates 
the performance of RFJAS through a direct, side-by-side comparison with 
WFS. We present a detailed noise budget for both techniques, with 
particular emphasis on limitations at low frequencies, below 30\,Hz. 
Our results show that WFS performance is constrained by technical noise 
arising from beam spot motion (BSM), mainly due to beam miscentering on QPDs. In contrast, RFJAS is primarily limited 
by residual RF amplitude modulation. A blended scheme that combines both 
sensing methods may offer the most practical approach for use 
in gravitational wave detectors such as Advanced LIGO.

\end{abstract}

\maketitle



\section{\label{sec:intro}Introduction}

Gravitational wave detectors such as Advanced LIGO (aLIGO)~\cite{aLIGO}, Advanced Virgo~\cite{AdVirgo}, and KAGRA~\cite{KAGRA}, 
require precise active control of their suspended optical 
cavities to achieve optimal interferometer performance. 
Several techniques exist for this purpose, with the currently
employed method being wavefront sensing (WFS)~\cite{Morrison, Morrison2}. 
In this approach, misalignments between optical cavities (or between an input beam and an optical cavity) are sensed using pairs of quadrant
photodetectors (QPDs) demodulated at the Pound-Drever-Hall (PDH)~\cite{PDH} phase modulation frequency. 
The QPDs detect beat signals between the $HG_{00}$ mode carrier and $HG_{10/01}$ mode PDH sidebands as well as between the $HG_{10/01}$ mode carrier and $HG_{00}$ mode sidebands. 
To optimally recover alignment information for a cavity interface the two QPDs must be placed with a Gouy phase separation of $90^\circ$~\cite{Morrison}. 

Given that alignment noise has persistently limited the LIGO interferometer performance at low frequencies ($<30$\,Hz)~\cite{PhysRevD.102.062003, O4_sensitivity}, 
alternative sensing schemes are being explored to improve sensitivity in 
this band. To this end, we have developed a technique called Radio Frequency Jitter Alignment Sensing
(RFJAS) and performed a comparative study against WFS. RFJAS uses an 
electro-optic beam deflector (EOBD) to 
generate RF sidebands in the first-order spatial $\text{HG}_{10/01}$ modes. 
These sidebands are offset from the carrier frequency by the higher-order mode (HOM) spacing frequency of a downstream optical cavity~\cite{Fulda}. 
The RFJAS method generates a diagonal sensing matrix following in-phase and 
quadrature-phase ($I\&Q$) demodulation of a beat 
between $HG_{10/01}$ carrier and $HG_{10/01}$ sidebands,
as detected by a single-element photodetector in reflection of the cavity.

\section{Theory} \label{Theory}
\subsection{Radio Frequency Jitter Alignment Sensing (RFJAS)}
The RFJAS scheme relies on rapid 
angular deflection of the laser beam. In our case, we are using an 
EOBD to achieve this. Jittering the laser
beam generates phase modulated sidebands in the first HOM. The EOBD is aligned to deflect in the x-direction 
that is parallel to the table plane, hence generating $HG_{10}$ sidebands.

\begin{figure}
    \caption{Alignment degrees of freedom (DOFs) of an cavity. On the top is a tilt DOF, and the bottom is the translation DOF. The black line is the cavity eigenaxis, while the red line is the misaligned input beam.}
    \includegraphics[width=0.4\textwidth]{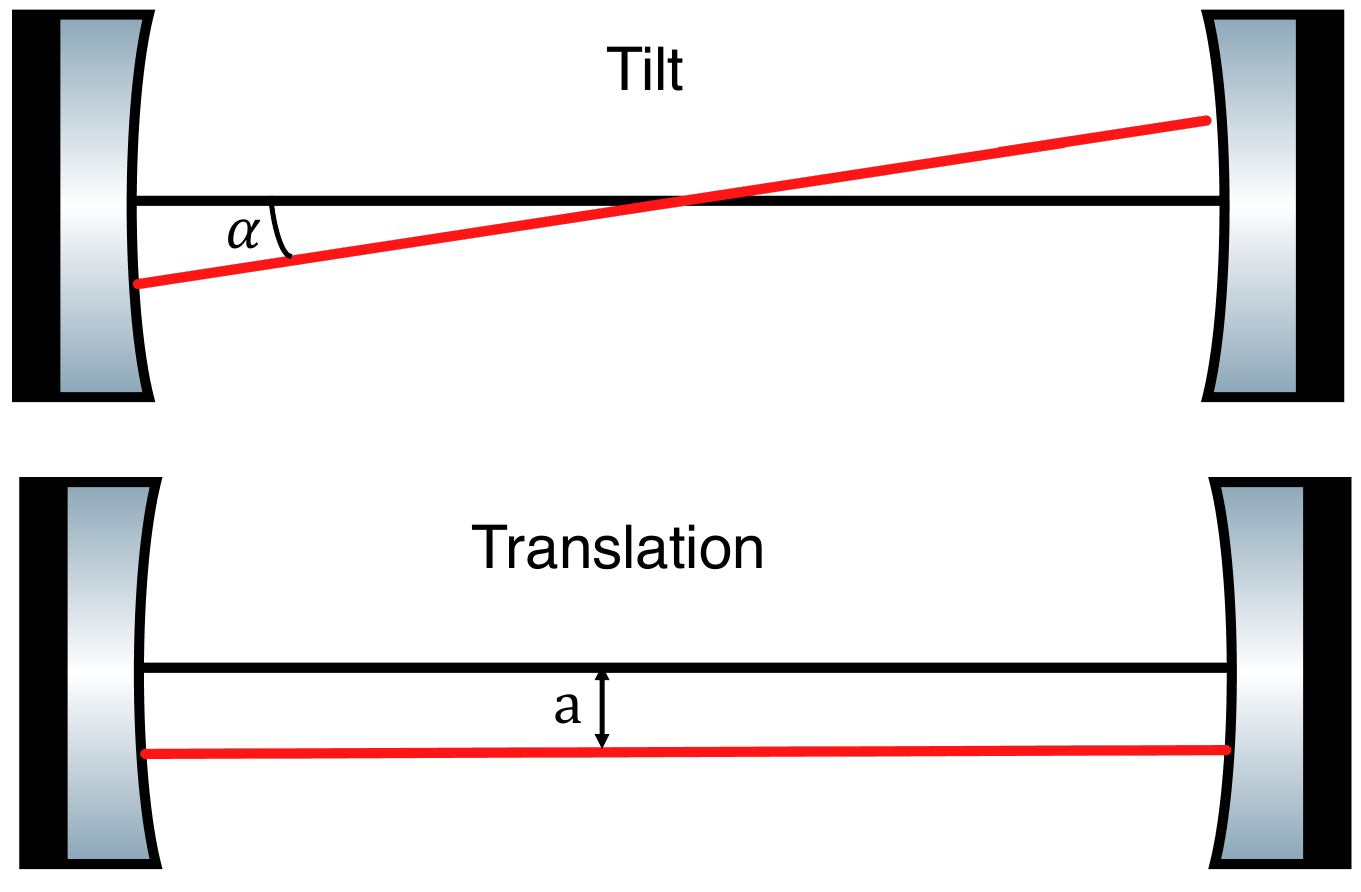}
    \centering
\label{cavity DOFs}
\end{figure}

The theory behind this scheme has been derived and published 
in Refs.~\cite{Fulda} and \cite{Liu}. Error signals were derived for tilt and translation misalignment
degrees of freedom, showing that in the case of an impedance matched 
cavity, a tilt misalignment by angle $\alpha$ produces a voltage from 
the photodetector on reflection 

\begin{equation}
    V^{\Omega,I} \approx \frac{E_{0}^2 m_{\alpha} \alpha }{\Theta_m \Theta_{cav}},
    \label{voltage Omega I}
\end{equation}
where $\Omega$ is the driving frequency of the EOBD and it is equal 
to the HOM separation frequency of the cavity, $I$ signifies the in-phase demodulation quadrature, $E_0$ is the field amplitude at the photodetector, $m_{\alpha}$ is the angular modulation amplitude, and $\Theta_m$ and $\Theta_c$ are the far-field divergence angles of the beam at the EOBD and the cavity waist, respectively. These divergence angles are given by
\begin{equation}
\begin{aligned}
    \Theta_{mod} &= \frac{\lambda}{\pi w_{0mod}} \quad
    &\Theta_{cav} &= \frac{\lambda}{\pi w_{0cav}}
\end{aligned}
\end{equation}

where $w_{0mod}$ is the beam waist size inside the EOBD.
In case of a lateral offset of amount $a$ the voltage produced by a quadrature-phase demodulation of the photodetector in reflection is 

\begin{equation}
    V^{\Omega,Q}\approx -\frac{E_{0}^2 m_{\alpha} a }{\Theta_m w_{0cav}}
    \label{voltage Omega Q}
\end{equation}

Equations \ref{voltage Omega I} and \ref{voltage Omega Q} show a linear 
relationship between the voltage readout from the orthogonal demodulation 
channels I and Q of the reflection photodetector and the key alignment degrees of freedom (DOFs). 

Figure~\ref{cavity DOFs} show the tilt and lateral alignment DOF in a single cavity. The black line is the eigenaxis of the cavity, and the red line is the misaligned input beam. 
\subsection{WaveFront Sensing (WFS)}








The conventional WFS scheme takes its name from its technique of detecting the phase difference between two wavefronts incident on a QPD~\cite{Morrison}. The technique is best understood however through the HOM formalism, which we will predominantly use here.
This scheme uses an Electro-Optic Modulator (EOM) 
to generate RF sidebands in the fundamental mode, which beat with any 
first-order HOM on the QPD in reflection of the cavity to 
detect misalignment.

Tao \emph{et al}.~\cite{Liu} derived the equations for both tilt and translation 
misalignment DOFs for an impedance-matched cavity. For a QPD that is placed at \(180^\circ \) of accumulated Gouy phase from the cavity waist, a tilt misalignment produces a voltage after in-phase demodulation that is linearly proportional to \(\alpha\):

\begin{equation}
    V_{\mathrm{WFS}}^{\mathrm{QPD}_1} \approx  -2 E_0^2 \frac{\alpha}{\Theta_{\mathrm{cav}}} m,
    \label{equ:WFS tilt}
\end{equation}

where $m$ is the phase modulation depth of the EOM, $\alpha$ is the 
angular misalignment, and $\Theta_{\mathrm{cav}}$ is the far-field divergence angle of the cavity.

For a second QPD that is \(90^\circ \) of accumulated Gouy phase from the cavity waist, a lateral misalignment produces a voltage after in-phase demodulation that is linearly proportional to \(a\) :

\begin{equation}
    V_{\mathrm{WFS}}^{\mathrm{QPD}_2} \approx  2 E_0^2 \frac{a}{w_{0\mathrm{cav}}} m,
\end{equation}

where $a$ is the amount of lateral displacement, and $w_{0\mathrm{cav}}$ is the waist size of the cavity.

\section{Experimental layout and Characterizations} \label{sec: 3}

\subsection{Experimental layout}
\begin{figure}  
\centering
\includegraphics[height=7cm]{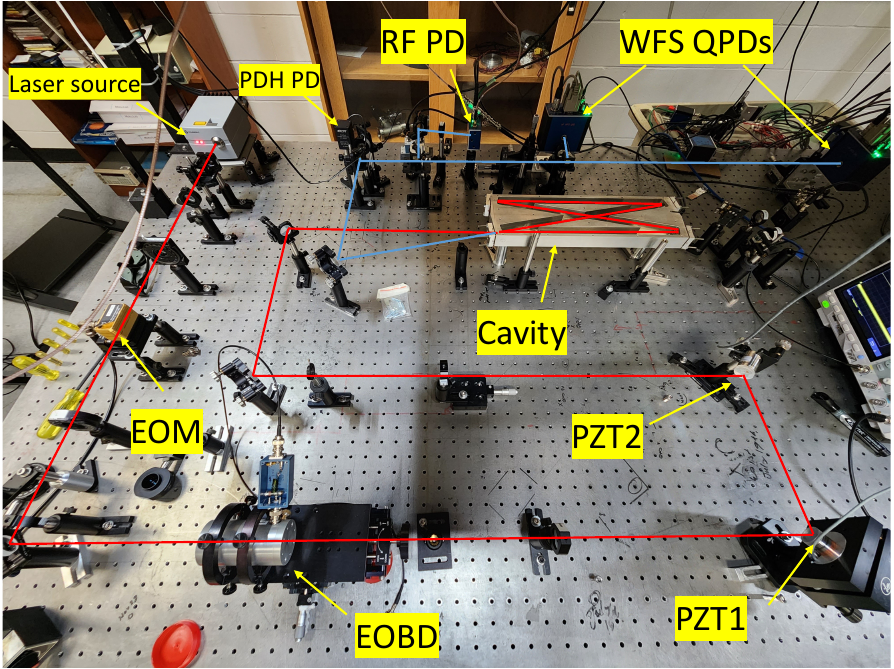}
\caption{The tabletop optical layout used in this experiment.}
\label{fig:Table setup}
\end{figure}

Figure~\ref{fig:Table setup} shows the schematic of the table layout. 
The setup begins with a laser source emitting light at 1064~nm, followed 
by an EOM used to generate fundamental mode 
sidebands for both Pound-Drever-Hall (PDH) locking and WFS alignment. 
Next in the beam path is the EOBD, generating HG$_{10}$ mode sidebands for RFJAS, followed by two steering mirrors labeled PZT1 and 
PZT2. These mirrors are separated by $90^\circ$ of 
Gouy phase, allowing actuation of orthogonal alignment degrees of 
freedom before the beam enters the bowtie optical cavity.

In reflection from the cavity, the setup includes a PDH photodetector for 
frequency locking, a  single-element radio-frequency 
photodetector (RFPD) used for the RFJAS 
scheme, and two QPDs with associated Gouy phase 
telescopes. It is worth noting that both RFJAS and PDH schemes can use the same photodetector on reflection, however, in our setup we use two separate photodetectors solely to have full control of both techniques separately, if troubleshooting was necessary.

\subsection{Characterization}

\subsubsection{Bowtie Cavity}
The cavity used in this setup is a bowtie cavity. It has four mirrors, the input and output couplers are flat and the other two mirrors are curved. The cavity has the following characteristics shown in Tab.~\ref{tab:Cavity parameters lab}.
This design is chosen to produce an astigmatic eigenmode; that is it has different waist sizes in the planes parallel and orthogonal to the table. This results in different HOM separation frequencies in both planes, allowing one to recover non-degenerate alignment signals at the RFPD for misalignments in both planes.

\begin{table}[ht]
    \centering
    \caption{Measured parameters of the bowtie cavity used in the setup.}
    \label{tab:Cavity parameters lab}
    \begin{tabular}{lcc}
    \toprule
    \textbf{Parameter} & \textbf{Symbol} & \textbf{Value} \\
    \midrule
    Round-trip optical path length & $L_{\mathrm{rt}}$ &  {1.6 m} \\
    Finesse & $\mathcal{F}$ & 990 \\
    Free Spectral Range & $\nu_{\mathrm{FSR}}$ &  {186 MHz} \\
    Full Width at Half Maximum & $\Delta \nu$ &  {184 kHz} \\
    Round trip Gouy phase & $\Delta \Phi_{\mathrm{rt}}$ & $66^\circ$ \\
    Curved Mirrors Radius of Curvature & RoC & {5 m} \\
    HOM Separation Frequency (x-axis) & \( \delta f_{HOM}^x\) & {34.74 MHz} \\
    HOM Separation Frequency (y-axis) & \( \delta f_{HOM}^y\) & {34.44 MHz} \\
    \bottomrule
    \end{tabular}
\end{table}

\subsubsection{Alignment DOF Driving Matrix}

In our setup PZT1 and PZT2 are placed with \(90^\circ\) of Gouy phase separation. 
This means that the PZT mirrors are orthogonal in terms of their alignment actuation in the cavity basis. It does not mean, however, that one PZT produces a tilt in the cavity basis while the other produces a translation; in general each one produces some combination of both tilt and translation.
Let us call the combination of tilt and translation in the cavity basis produced by PZT1 \(C1\), and the equivalent PZT2 combination \(C2\). The driving matrix then is shown in Tab.~\ref{alignment DOF drive matrix}.

\begin{table}[htb]
\begin{center}
    \begin{tabular}{ |c|c|c| } 
     \hline
      PZTs/Combinations & C1 & C2 \\ 
      \hline
      PZT1 & 1 & 0 \\ 
      \hline
      PZT2 & 0 & 1 \\ 
     \hline 
    \end{tabular}
    \caption{Driving matrix of the alignment DOFs.}
    \label{alignment DOF drive matrix}
\end{center}
\end{table}

The exact combinations of PZT mirror drives that result in the cavity basis tilt and translation DOFs are not known. Instead, we describe the sensing schemes in terms of the orthogonal basis of the PZT mirrors themselves.

\subsubsection{EOBD Modulation Frequency Sweep Measurement}

This measurement demonstrates the effectiveness of the RFJAS scheme by 
generating clearly separated alignment signals for our optical cavity. 
To achieve this, we drive the alignment degrees of freedom using PZT1 and 
PZT2, following the drive matrix in Tab.~\ref{alignment DOF drive matrix}, 
while sweeping the EOBD modulation frequency across the cavity’s 
HOM separation frequency, \( \delta f_{HOM}^x\). At each frequency point, 
we collect the in-phase and quadrature-phase (I\&Q) signals and construct 
a sensing matrix, as shown in Tab.~\ref{tab:sensingmatrixRFJ}.

\begin{table}[htb]
\begin{center}
    \begin{tabular}{ |c|c|c| } 
     \hline
     PZTs/RFJAS & I & Q \\ 
     \hline
     PZT1 & $I_{PZT1}$ & $Q_{PZT1}$ \\ 
     \hline
     PZT2 & $I_{PZT2}$ & $Q_{PZT2}$ \\ 
     \hline 
    \end{tabular}
\end{center}
\caption{Sensing matrix for RFJAS response to the PZTs driving matrix.}
\label{tab:sensingmatrixRFJ}
\end{table}
For each matrix, we define the complex signals:
\begin{equation}
\begin{split}
    \widetilde{PZT1} &= I_{PZT1} + iQ_{PZT1} \\
    \widetilde{PZT2} &= I_{PZT2} + iQ_{PZT2}
\end{split}
\end{equation}
We then apply a demodulation phase rotation to both complex quantities 
by the phase:
\begin{equation}
\phi=\arctan(Q_{PZT1}/I_{PZT1})
\end{equation}
yielding the rotated components:
\begin{equation}
\begin{split}
    I_{PZT1}' &= \mathscr{Re}(\widetilde{PZT1} \cdot e^{-i\phi}) \\
    Q_{PZT1}' &= \mathscr{Im}(\widetilde{PZT1} \cdot e^{-i\phi}) \\
    I_{PZT2}' &= \mathscr{Re}(\widetilde{PZT2} \cdot e^{-i\phi}) \\
    Q_{PZT2}' &= \mathscr{Im}(\widetilde{PZT2} \cdot e^{-i\phi}).
\end{split}
\label{eq:MatrixRotation}
\end{equation}

This rotation ensures that $Q_{PZT1}'$ is forced to zero at all frequencies, 
and that $I_{PZT2}'$ goes to zero at the cavity’s HOM separation frequency.
Alternative rotation methods are possible, which should also result in a diagonal sensing matrix at the HOM frequency.

\begin{figure}[ht]
\centering
\includegraphics[width=\linewidth]{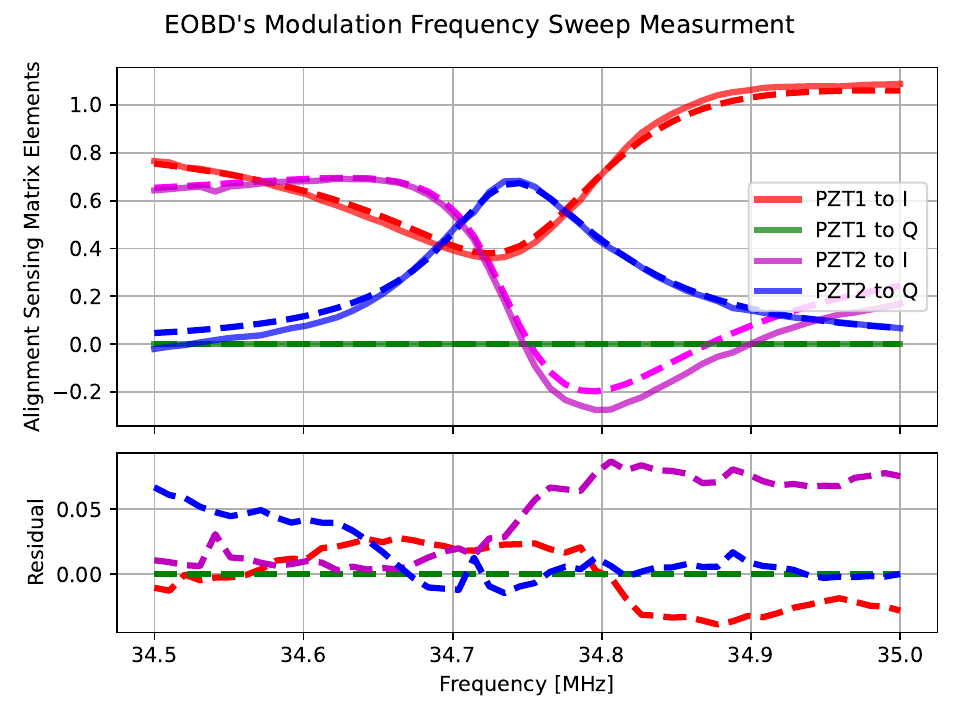}
\caption{Alignment sensing matrix after phase rotation as a function of modulation frequency. The solid lines are the measured data and the dashed lines are the simulation fit data. The plots are in units of V/V (data) or W/W (simulation), as it is normalized to the quadrature sum of the first sensing matrix elements. The asymmetry in the shape of the PZT2 to Q curve and PZT1 to I curve is due to the accumulated Gouy phase between the EOBD and the cavity waist. Different Gouy phases correspond to different shapes of the sweep measurement. However, the reason that PZT1 to I does not have the same value as PZT2 to Q at the zero-crossing of PZT2 to I could be due to the fact the PZTs do not drive the same misalignment DOF amplitude, as well as the presence of mode mismatch.}
\label{fig:frequency sweep}
\end{figure}

Figure~\ref{fig:frequency sweep} shows the resulting sensing matrices as a 
function of modulation frequency after the phase rotation. The PZT2 component 
in the in-phase demodulation channel crosses zero precisely at the HOM frequency, 
indicating a diagonal sensing matrix. 
This confirms that the in-phase demodulated signal 
primarily senses alignment of PZT2, while the quadrature-phase demodulation signal primarily senses alignment of PZT1. 

We simulate this measurement using  FINESSE3~\cite{finesse} with the same setup parameters as the table-top experiment, 
and observe good agreement between the experimental data and the simulation, as shown in Fig.~\ref{fig:frequency sweep}. 
The plot is 
normalized to the quadrature sum of all elements in the first sensing matrix for 
both the simulation and experimental data.

\subsubsection{Wavefront Sensing (WFS)} \label{sec:radarplot}

For the WFS scheme, we use two QPDs to detect 
alignment signals in reflection of the cavity. The beam size on QPD1 is 
approximately 430\,$\upmu$m, and on QPD2 it is around 470\,$\upmu$m. 
The beam waist between them is about 311\,$\upmu$m. 
This geometry places the QPDs at a relative Gouy phase difference of roughly $90^\circ$.

\begin{figure}[ht]
    \centering
    \includegraphics[width=0.7\linewidth]{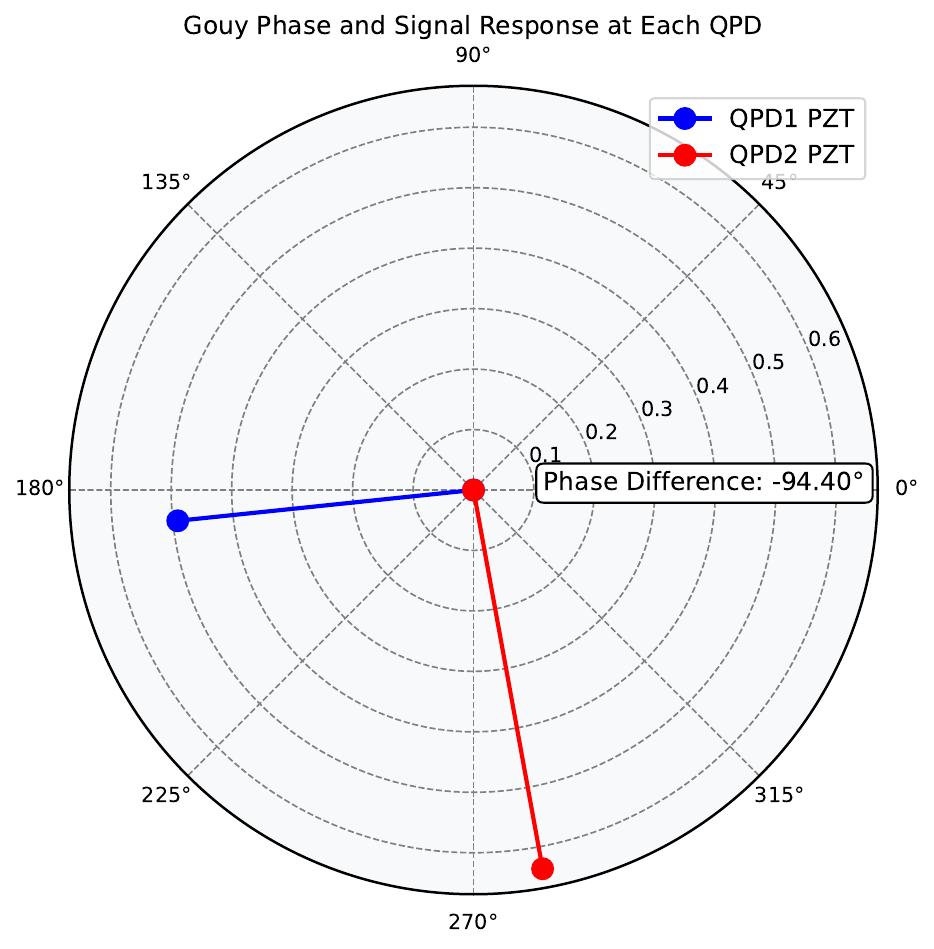}
    \caption{Polar representation of the Gouy phase separation between QPD1 and QPD2. The radial component magnitude corresponds to the magnitude response of each QPD to the driving alignment DOF in volts. }
    \label{fig:gouy phase QPDs}
\end{figure}

We detect alignment error signals that are linear combinations of the 
responses from PZT1 and PZT2. These signals populate the WFS sensing matrix shown in Tab.~\ref{QPDs response to PZTs drive}:

\begin{table}[!h]
\begin{center}
\begin{tabular}{ |c|c|c| } 
 \hline
  PZTs/QPDs & QPD1 [V] & QPD2 [V] \\ 
 \hline
  PZT1 & -0.489 & 0.114 \\
 \hline
  PZT2 & -0.050 & -0.626 \\
 \hline
\end{tabular}
\end{center}
\caption{The WFS sensing matrix to driving the PZTs steering mirrors}
\label{QPDs response to PZTs drive}
\end{table}
We express the QPD responses to the PZT drives as complex numbers in the form:
\begin{equation}
    \begin{split}
        \widetilde{QPD1_{PZT}}= -0.489 -0.050i \\
        \widetilde{QPD2_{PZT}} = 0.114 -0.626i
    \end{split}
    \label{QPD rotation}
\end{equation}
Similar to the RFJAS case, we apply a complex phase rotation to this 
matrix to diagonalize the sensing response.
Moreover, the angle of each response in the complex plane can be calculated as
\begin{equation*}
    \theta_{QPD} = \arctan\!\left(\frac{c}{a}\right),
\end{equation*}
 where $a$ and $c$ are the real and imaginary components of each QPD response from Eqn~\ref{QPD rotation}, respectively. This angle implies the response of the QPD to each alignment DOF in the PZT mirrors' actuation basis. 
Since we drive orthogonal alignment degrees of freedom, we can plot the response of both QPDs and confirm their orthogonality on a polar plot, as seen in Fig.~\ref{fig:gouy phase QPDs}. The angle difference, \(\theta_{QPD1} - \theta_{QPD2}\), represents the Gouy phase difference between QPDs, and the radial component is the magnitude of the response in volts. From the plot we can see that the Gouy phase separation of the QPDs is 94.4$^\circ$, which is acceptably close to the optimal 90$^\circ$ for the purposes of this experiment.

\subsubsection{Shot Noise} \label{sec: shot nosie theory}
Although shot noise is not a limiting factor in this experiment, it is a 
fundamental noise source in gravitational wave detectors such as LIGO. 
Shot noise originates from the quantum nature of light: photons arrive 
at the photodetectors according to a Poisson process, leading to statistical 
fluctuations in the measured signal \cite{Caves}. These fluctuations set 
a fundamental limit on the precision of phase measurements in the 
interferometer.

The shot noise 
scales with the square root of the detected photon number, while the signal scales linearly with the detected photon number, meaning that 
higher optical power reduces relative fluctuations.

The amplitude spectral density of shot noise depends on the incident 
power \( P_{\text{DC}} \) and photon energy \( \hbar \omega \), where 
\( \omega = 2\pi c/\lambda \), \(c\) is the speed of light and \(\lambda\) is the laser's wavelength:
\begin{equation}
    \sqrt{S_{\text{shot noise}}(f)} = \sqrt{2 \hbar \omega P_{\text{DC}}} 
    \quad [\mathrm{W}/\sqrt{\mathrm{Hz}}]
\label{eq:shot_nosie}
\end{equation}



For an impedance matched cavity held on resonance for the carrier HG$_{00}$ mode, the beam in reflection is mostly composed of the PDH phase modulation sidebands, along with RFJAS sidebands. If the PDH sideband power is the dominant contributor to $P_{DC}$, the shot noise will be largely independent of the RFJAS modulation depth. The RFJAS signal response to misalignments, on the other hand, scales linearly with RFJAS modulation depth. We find therefore that the signal-to-shot noise ratio of the RFJAS scheme will be proportional to the RFJAS modulation depth. 

The EOBD device provides a deflection gain on the order of 4\,$\upmu$rad/V and has an aperture that allows beams of waist size up to around 1\,mm, and therefore far-field divergence angles around 340\,$\upmu$rad. Assuming a 10\,V amplitude drive signal, the ratio $m_\alpha/\Theta_m$ therefore has a value of around 0.12, generating RFJAS sidebands with a combined power of $\frac{1}{2}(m_\alpha/\Theta_m)^2$, or around 0.7\,\% of the carrier power. To put this in context, the phase modulation depths used in aLIGO are 0.18 for both the 9\,MHz and 45\,MHz modulation frequencies, resulting in total PDH sideband power equal to 3.3\,\% of the carrier power at the interferometer input. Comparing Eqns.~\ref{voltage Omega I} and~\ref{equ:WFS tilt}, we see that the signal responses for RJFAS and WFS differ only by a factor 2, and the relative modulation depths.

We see therefore that the shot noise limited performance of the RFJAS scheme can be expected to be around the same level as that of the WFS scheme, especially if higher EOBD drive amplitudes can be achieved. Given that the RFJAS sensor could be the same sensor as the PDH sensor, the RFJAS scheme may also benefit from the shot noise advantage of a higher light power than the QPDs used for WFS. Next, we explore other noise couplings that impact the two alignment sensing schemes in more distinct ways.



\section{Results} \label{sec: 4 results}

In this section we report on the results of our comparison between both schemes. We start with discussing the beam spot motion (BSM) problem on both schemes. Then, we move to discuss the full noise budget of both schemes. 

\subsection{Beam Spot Motion (BSM) Coupling Measurement} \label{sec:BSM}
\subsubsection{WFS}

QPDs require the beam to be well centered between 
all quadrants. If the beam drifts, e.g. due to seismic motion, spurious 
alignment signals appear. To study this effect, we look at how BSM couples to the yaw signal on the QPDs.

\begin{figure}[ht]
\centering
{\includegraphics[height=0.45\linewidth]{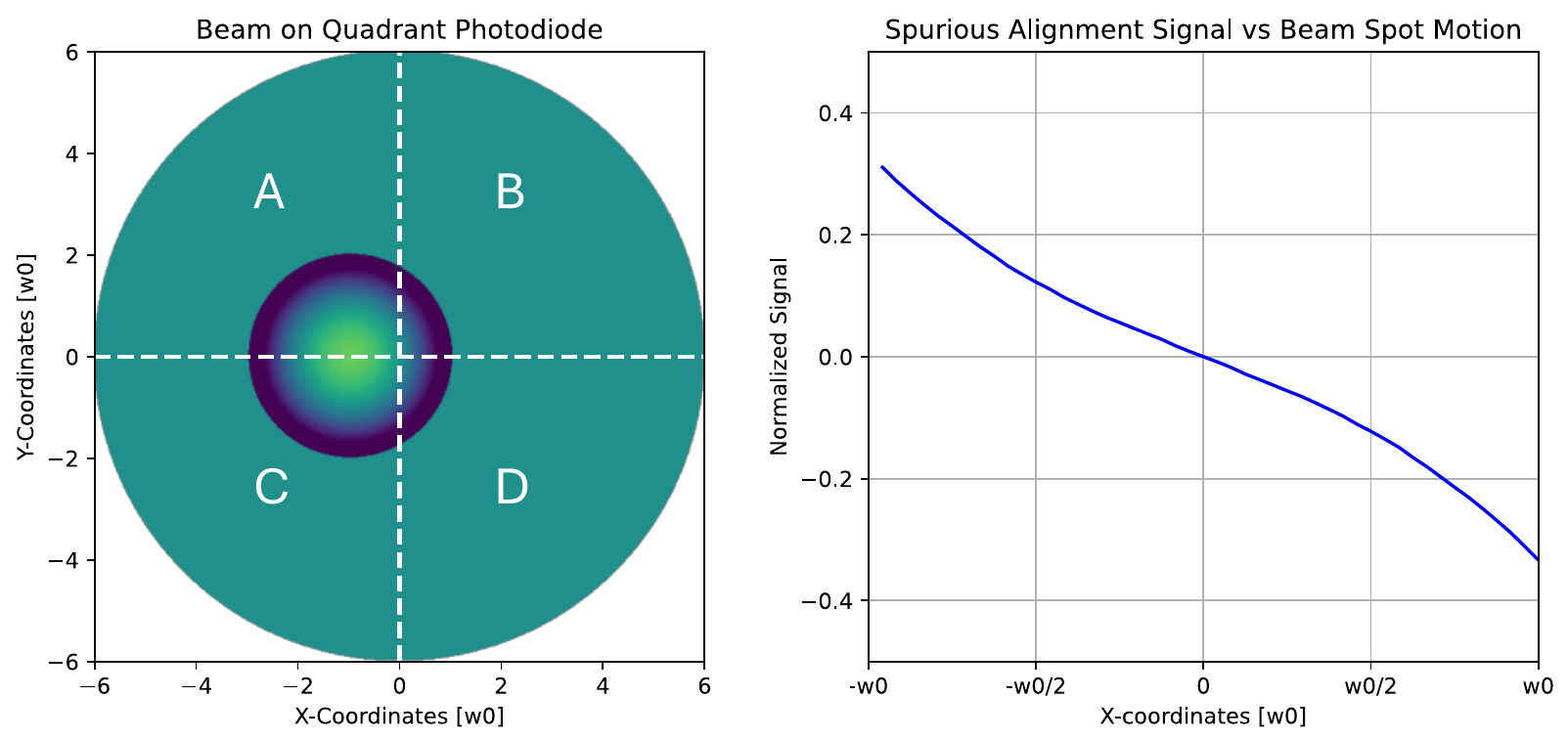}}
  \caption{Left: A visualization of the wavefronts shifted to the left and right of the center of the QPD by one beam radius. Right: The difference between RF demodulated signals from the left and rights halves, normalized by the sum of the DC power on both halves, as a function of spot position offset.}
\label{fig:screenshot}
\end{figure}

We start with the cavity fully aligned, mode matched, and the beam centered 
on the QPDs. In this condition, the RF and DC yaw readouts, calculated as 
$A+C-B-D$ in Fig.~\ref{fig:screenshot}, are zero. Using motorized 
micrometers (ULN2003 stepper motors), we translate each QPD in $31.25~\upmu$m 
steps, up to a total of $250~\upmu$m to the right, recenter, and then repeat 
to the left. At each step, we record the yaw signal and calculate the slope 
around the center. We repeat this for six different mode mismatches: 
$0.6\%$, $2.2\%$, $4.5\%$, $5.7\%$, $7.7\%$, and $11.1\%$.

When the cavity is mode matched, the resonant carrier and non-resonant sidebands have the same wavefront curvature as each other in reflection. When a mode mismatch is present, this may not be the case. In the HOM picture, the mismatch couples some power into the second-order HOMs ~\cite{Anderson_technique}. Interference between the 
fundamental mode and these HOMs produces a symmetric phase pattern that 
cancels out when the beam is centered on the QPD. As soon as the beam is offset, however, this symmetry breaks and a spurious yaw signal appears. 
Increasing the mode mismatch increases the curvature of the HOM wavefront, which in turn makes the signal more sensitive to miscentering.

Figure~\ref{fig:screenshot} shows an example of the spurious alignment signal appearing from the combination of mode mismatch and BSM. On the left is the phase 
difference between a mismatched and a mode-matched beam, displaced from the QPD center (color indicates the phase difference). 
On the right, we plot the corresponding spurious alignment signal: the difference between QPD's left and right side RF demodulated outputs, with the spot position offset range given in units of the beam radius at the QPD. The RF demodulated output is normalized by the total DC power on the QPD.


\begin{figure}[ht]
    \centering
    {\includegraphics[width=0.9\linewidth]{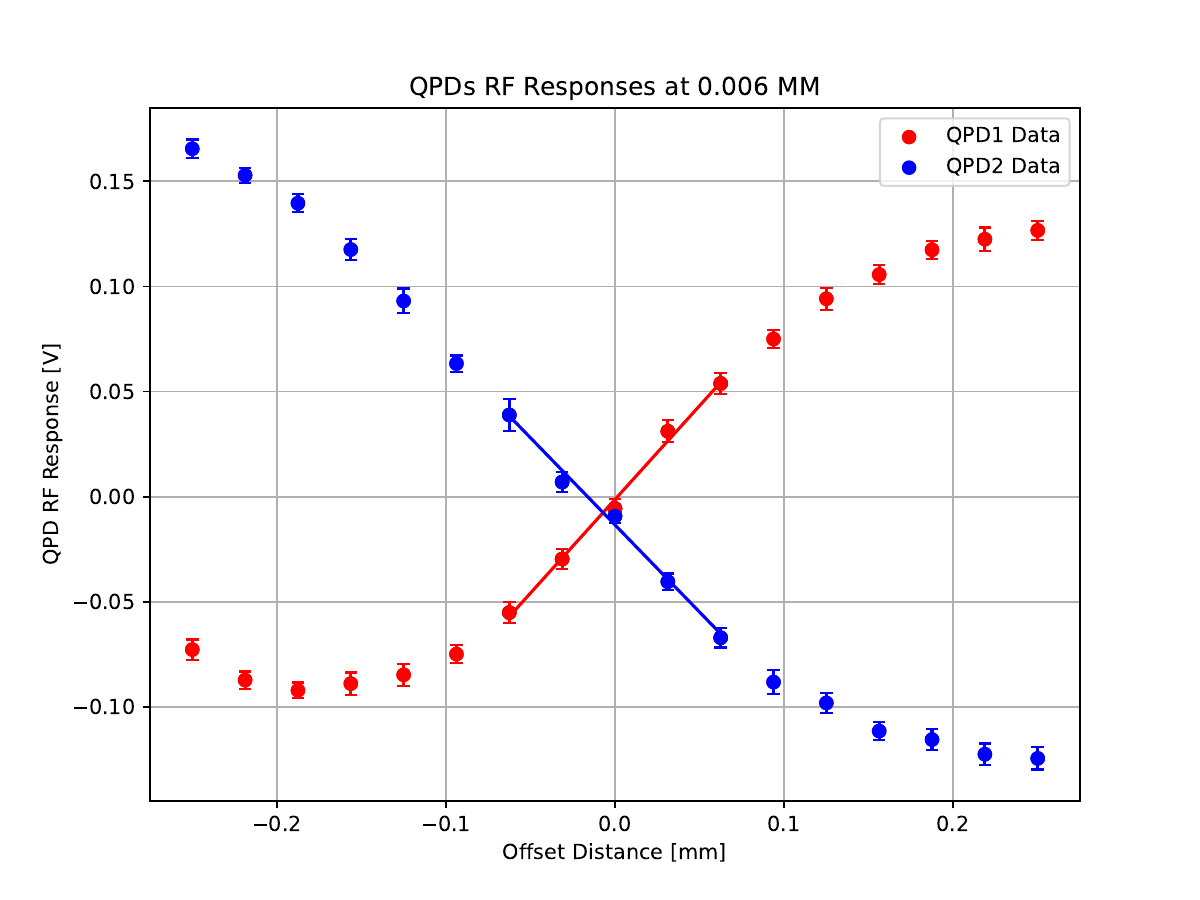}}
    \caption{The RF signals on both QPDs as a function 
    of translation distance at the lowest mode mismatched achieved in the experiment, 0.6$\%$. On the x-axis is the distance both QPDs translated, in mm, while on the y-axis is the voltage coming from the RF channels with error bars on them.}
    \label{fig:RF BSM at 0.6 MM}
\end{figure}

\begin{figure}[ht]
    \centering
    {\includegraphics[width=0.9\linewidth]{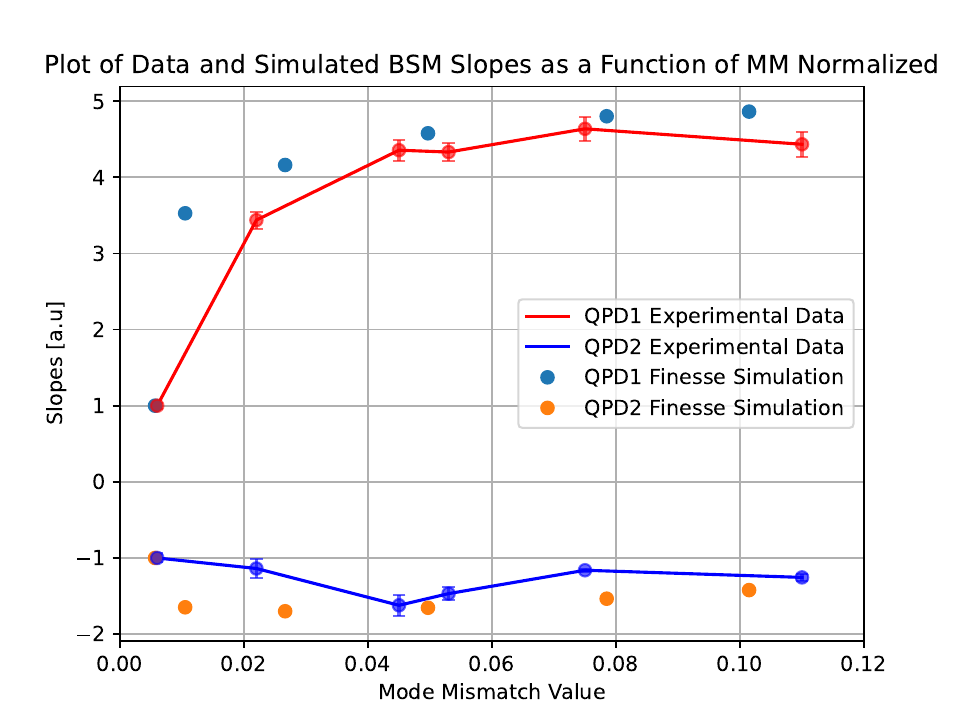}}
    \caption{A plot of the experimental and the simulated slopes of the RF responses of both QPDs as a function of mode mismatch. On the x-axis is the mismatch value at which each data set is taken, and y-axis is the slope of each set. The slopes are normalized to the slope of the lowest mismatch}
    \label{fig:beam spot slopes}
\end{figure}

In the case of the lowest mismatch achieved in the experiment, \(0.6 \%\),
both QPDs have almost the same spurious signal amplitude and
slope, as seen in Fig.~\ref{fig:RF BSM at 0.6 MM}. 
As the mismatch 
increases, the beam size increases on QPD1 and decreases 
on QPD2. This will result in different 
RF and DC responses to BSM. 
The calculated slopes are plotted against the power mismatch value, in Fig.~\ref{fig:beam spot slopes}. 

This effect can also be simulated using FINESSE3. A setup similar to the tabletop 
setup is built in the simulation with similar Gouy phases, beam sizes, and mode matching. The qualitative behavior in the simulation is similar to the experimental data. The slopes of the FINESSE simulation are also plotted in Fig. ~\ref{fig:beam spot slopes} for comparison.

Figure \ref{fig:beam spot slopes}
is normalized to the first mode mismatch slope value. These measurements of the BSM to spurious alignment signal coupling, and their good agreement with the model, allow us to form a term for BSM noise in the WFS noise budget.

The difference in slopes and response of both QPDs to BSM not only depends on the beam size and the mismatch magnitude, but also on the mismatch type; i.e. waist size mismatch or waist position mismatch.


\subsubsection{RFJAS}
The RFJAS scheme uses a single-element photodetector instead of QPDs, so in theory, beam miscentering should have no effect on the alignment signals. This constitutes a clear advantage over the classical WFS, which is sensitive to beam spot motion. In practice, however, the finite size of the beam and the active area of the photodetector can introduce some dependence of the signal on beam position. We therefore allow for a small contribution of beam spot motion in the RFJAS noise budget, while noting that this is a practical, rather than fundamental, limitation of the scheme.


\subsection{Noise budget} \label{sec: noise budget}

A noise budget for both schemes was compiled. 
The noise budget describes
the effect of various noises on the two schemes' sensing capability of misalignments, over a range of frequencies
roughly corresponding to the ground-based gravitational wave detection band. The noise is expressed in terms of \(\frac{|HG_{10}/HG_{00}|}{\sqrt{Hz}}\), i.e. the relative first order mode amplitude per \(\sqrt{Hz}\), following the formulation used by Mueller in Ref.~\cite{Guidojitter}. Furthermore, since the responses of QPD1 and QPD2, as well as I-channel and Q-channel, are similar in behavior and magnitude, each noise source is expressed as the quadrature sum of the noises in orthogonal sensing channels as follows:

\begin{equation*}
    WFS_{noise}=\sqrt{QPD1_{noise}^2 +QPD2_{noise}^2  }
\end{equation*}
and similarly for RFJAS:
\begin{equation*}
    RFJAS_{noise}=\sqrt{I_{noise}^2 +Q_{noise}^2  }
\end{equation*}

Our noise budget includes the electronics noise, residual RF amplitude modulation (RFAM), and BSM noise, projected from the measurements in section~\ref{sec:BSM}. The theoretical
shot noise is also calculated as in section \ref{sec: shot nosie theory} but not plotted as it is constant in frequency, and well below all other noises. The total noise, which is the quadrature sum of all noise budget terms, is also plotted, along with the actual amplitude spectral density of the alignment sensing channels in full operation.

\begin{figure*}[ht]
    \centering
    {\includegraphics[width=\linewidth]{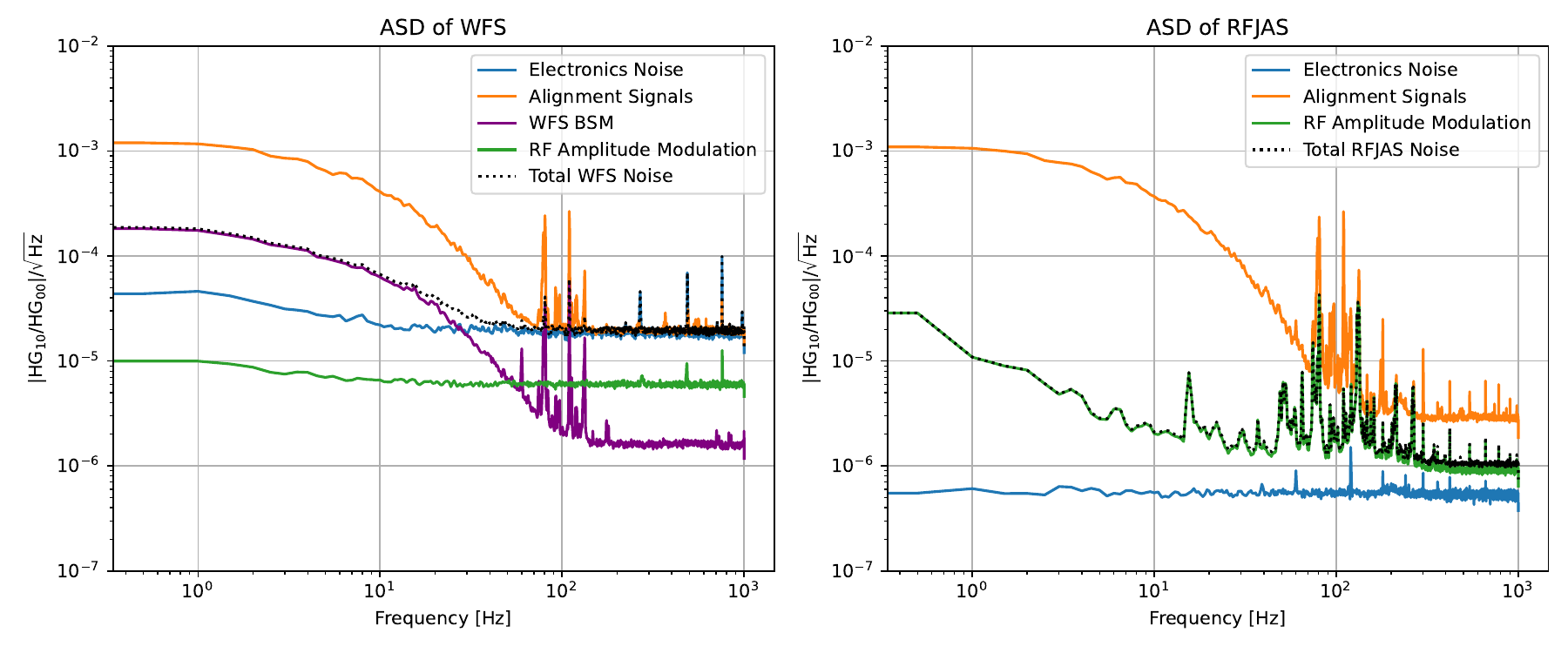}}
    \caption{The measured noise budget of both alignment sensing schemes. On the left is the noise spectra of the QPDs used for WFS, and on the right are the spectra of \(I\&Q\) channels of RFJAS. The y-axis is in units of \(\frac{|HG_{10}/HG_{00}|}{\sqrt{Hz}}\). On both plots are the electronics noise in blue color, RFAM noise in green, and quiescent alignment noise in orange. For the QPDs, BSM noise is plotted in purple. This is the dominant noise for WFS. While RFAM is dominant noise source for RFAS.}
    \label{fig:Noise budget}
\end{figure*}

\subsubsection{Calibration}\label{sec:cal}
For the noise budget, it is important to characterize the sensing matrices and calibrate the driving PZTs. Here we explain how we can convert from \(\frac{V}{\sqrt{Hz}}\) to \(\frac{|HG_{10}/HG_{00}|}{\sqrt{Hz}}\).

\subsubsection*{WFS}

Here we provide the diagonalized sensing matrices per driving voltage of the PZTs. These values will be used later to project the sensors' voltages to higher order mode content


\begin{center}
\begin{tabular}{ |c|c|c| } 
 \hline
  PZTs/QPDs & QPD1' [\(V_s\)] & QPD2' [\(V_s\)] \\ 
 \hline
  PZT1 [\(V_d\)] & 0.035 & 0.00092 \\
 \hline
  PZT2 [\(V_d\)] & 0 & 0.04557 \\
 \hline
\end{tabular}
\label{Diagonalized_QPDs_noise_sensing_matrix}
\end{center}
\(V_d\) is the driving voltage of the PZTs and \(V_s\) is the sensors response voltage to the PZTs after diagonalization.

\subsubsection*{RFJAS}
In a similar procedure, here is the diagonalized RFJAS sensing matrix


\begin{center}
\begin{tabular}{ |c|c|c| } 
 \hline
  PZTs/QPDs & I' [\(V_s\)] & Q' [\(V_s\)] \\ 
 \hline
  PZT1 [\(V_d\)] & 0.0042 & -0.000078 \\
 \hline
  PZT2 [\(V_d\)] & 0 & 0.003328 \\
 \hline
\end{tabular}
\label{Diagonalized_IQ_noise_sensing_matrix}
\end{center}
These diagonalized matrices will be used to calibrate the noise budget.

\subsubsection*{Calibrating the PZTs}
To characterize the PZTs, we calculate the driving angle per driving volt, \(V_d\). PZT1 drives \( \alpha_{PZT1} = 9\upmu rad / V_d\) and PZT2 drives 
\(\alpha_{PZT2} = 11\upmu rad /V_d\). This is measured by recording the deflection of the beam on a CMOS sensor 50 cm away from PZT1. The beam size on both PZTs is \(\approx 290 \upmu m\).

To calculate the higher order mode content, we use the approximation of a general tilt misalignment 

\begin{equation}
    HG_{00}^\alpha = HG_{00} e^{-i \Psi} + \frac{i\alpha}{\zeta} HG_{10} e^{-2i \Psi}
\end{equation}
where 
\begin{equation*}
    \zeta = \frac{2}{k w}
\end{equation*}
k is the wavenumber at \(1064 \times 10^{-9}\) \,m wavelength \(w\) is the beam size, and \(\Psi\) is the Gouy phase at the tilt location.

This means, at the PZTs, \(\zeta_{PZT} = 0.0011 rad\). The value of \(HG_{10}/HG_{00}\) is equal to \(\alpha/\zeta\)

\begin{equation*}
\begin{aligned}
\text{PZT1:}\quad & \frac{\alpha_{PZT1}}{\zeta_{PZT}} = 0.0079 \,\frac{|HG_{10}/HG_{00}|}{V_d}, \\[6pt]
\text{PZT2:}\quad & \frac{\alpha_{PZT1}}{\zeta_{PZT}} = 0.01001 \,\frac{|HG_{10}/HG_{00}|}{V_d}.
\end{aligned}
\end{equation*}

Finally, to project the sensors' amplitude spectral densities (ASDs) to higher order mode content, we need the following calibration 

\begin{equation}
    \frac{|HG_{10}/HG_{00}|}{\sqrt{Hz}} = \frac{|HG_{10}/HG_{00}|}{V_d} \times \frac{V_d}{V_s} \times \frac{V_s}{\sqrt{Hz}} 
\end{equation}

where

\begin{itemize}[leftmargin=*, itemsep=2pt, topsep=0pt]
  \item $\dfrac{V_s}{\sqrt{\mathrm{Hz}}}$ : calculated ASD of the sensors voltages
  \item $\dfrac{V_d}{V_s}$ : reciprocal of the diagonalized sensing matrix elements
  \item $\dfrac{|HG_{10}/HG_{00}|}{V_d}$ : HOM content calibration factor of the PZTs
\end{itemize}

These values are then used to project the noise spectra, after diagonalization, to \(\frac{|HG_{10}/HG_{00}|}{\sqrt{Hz}}\)

\subsubsection{Electronic noise}
This is the noise inherent in the electronics of the QPDs and the RF photodetector. 
We measure
this noise by turning off the laser and recording the electronics signals.
Results show that the electronics noise is the limiting
noise source for WFS above 30 Hz, while it is not limiting RFJAS.
Since WFS and RFJAS signals are being acquired with the same NI-9220 DAQ, 
ADC noise is eliminated as a possible source of the difference in the noise levels.

QPDs are designed with gaps between quadrants of a minimum size in order to minimize electronic cross-talk between quadrants. This leads to a design pressure towards larger active areas for the quadrants in order to maintain optical efficiency of detection by minimizing the fraction of light power incident on the gaps. Larger active areas generally come with the penalty of higher capacitance, which can lead to increased charge noise coupling in active transimpedance stage readout circuit designs. However, other readout circuit designs are possible, whose noise performance is less dependent on the diode capacitance. In principle the electronics noise from QPDs and single-element RFPDs can be made comparable, but the design effort required to achieve this for QPDs is considerably greater.

\subsubsection{RF Amplitude Modulation}
When employing an EOM or EOBD to apply phase or angular modulation to a beam, some amount of residual RF amplitude modulation (RFAM) will invariably be applied to the beam at the same modulation frequency. In the case of PDH sensing, RFAM at the PDH modulation frequency can cause a PDH locking offset~\cite{RFAM_suppression}. If the RFAM level is not stable, this can lead to noise in the frequency difference between the laser and the cavity resonance frequency. In a similar way, RFAM at the RFJAS modulation frequency can lead to spurious alignment signals in the demodulated RFPD outputs. In our experiment, the dominant RFAM source in the RFJAS scheme comes from polarization modulation at the EOBD, which arises from the mismatch between the input beam's 
polarization vector and the crystal's crystallographic axis. This polarization modulation becomes RFAM after the beam encounters polarizing optics downstream of the EOBD.



The RF demodulated QPD outputs in the WFS scheme are nominally insensitive to RFAM at the PDH modulation frequency, because the signals are produced from the subtraction of left from right (for horizontal motion) or top from bottom (for vertical motion). However, constant beam spot position offsets or even BSM at the QPDs can provide a coupling path for RFAM noise into the WFS QPD outputs. 

The RFAM noise at the RFPD was measured in the absence of true alignment signals by recording the I and Q demodulated RFPD outputs while the cavity was unlocked. The actual RFAM experienced in-lock corresponds to the measured unlocked-cavity RFAM multiplied by a suppression factor related to the impedance matching factor of the cavity~\cite{RFAM_suppression}, which for this experiment was determined to be 0.33. The suppression factor might be frequency-dependent in general, however, in our frequency range it was found to be not changing. The voltage noise on the RFPD demodulated outputs was scaled by this impedance matching factor, and projected to the RFJAS noise budget via the calibration factors shown in Section~\ref{sec:cal}.

A similar measurement was used to estimate the RFAM noise in the WFS scheme. However, a combination of the higher electronic noise level in the QPDs, and the common mode subtraction of RFAM in the differenced outputs prevented us from accurately measuring the RFAM noise contribution to the WFS scheme, as this noise was smaller than the electronics noise. We nonetheless projected the result to the WFS noise budget, as a representation of the upper limit of RFAM noise for that scheme. This upper limit term is effectively just the measured electronics noise term, but reduced by the impedance matching factor for the cavity.

\subsubsection{Beam Spot Motion Noise}

This is the noise of the spurious alignment signals as 
discussed in Section~\ref{sec:BSM}. To measure this noise, we 
record the DC output of the QPDs in the quiescent state. 
From the measured slope of the DC channels in response to the BSM, we get the beam motion on the QPDs. Multiplying 
the motion by the RF slope at \(0.6\%\) mismatch (seen in Fig.~\ref{fig:RF BSM at 0.6 MM}), gives the contribution of the BSM to the demodulated QPD output channels. This is the limiting noise source 
for WFS scheme up to around 30\,Hz. RFJAS does not suffer 
from this noise as it utilizes a single element photodetector.


\subsubsection{Shot Noise}
The theoretical shot noise is also calculated and it is constant in frequency found to be below all other noises. Therefore, it was not plotted in the noise budget. 
The calculations of the shot noise can be found in appendix ~\ref{appendix:shot_noise}. However, we include the final results here: 
The shot noise in the RFJAS scheme is 
\begin{equation}
\text{RFJAS}_{\text{SN}}= 2.71 \times 10^{-7} \,\frac{|HG_{10}/HG_{00}|}{\sqrt{\mathrm{Hz}}}
\end{equation}

while the shot noise in WFS is 

\begin{equation}
    S_{WFS}=\sqrt2 \times 1.6\times 10^{-7} \approx 2.26\times 10^{-7} \,\frac{|HG_{10}/HG_{00}|}{\sqrt{\mathrm{Hz}}}
\end{equation}

\subsubsection{Total Noise}
The total noise is calculated as the quadrature sum of the noise ASDs as follows

\begin{equation}
    N_{\text{total}} = \sqrt{\sum_{i=1}^{N} n_i^2}
\end{equation}
where $n_i$ is the ASD of the independent noise sources, which we assume to be are uncorrelated with each other. 

\subsection{Coherence}
\begin{figure}[ht]
    \centering
    \includegraphics[width=\linewidth]{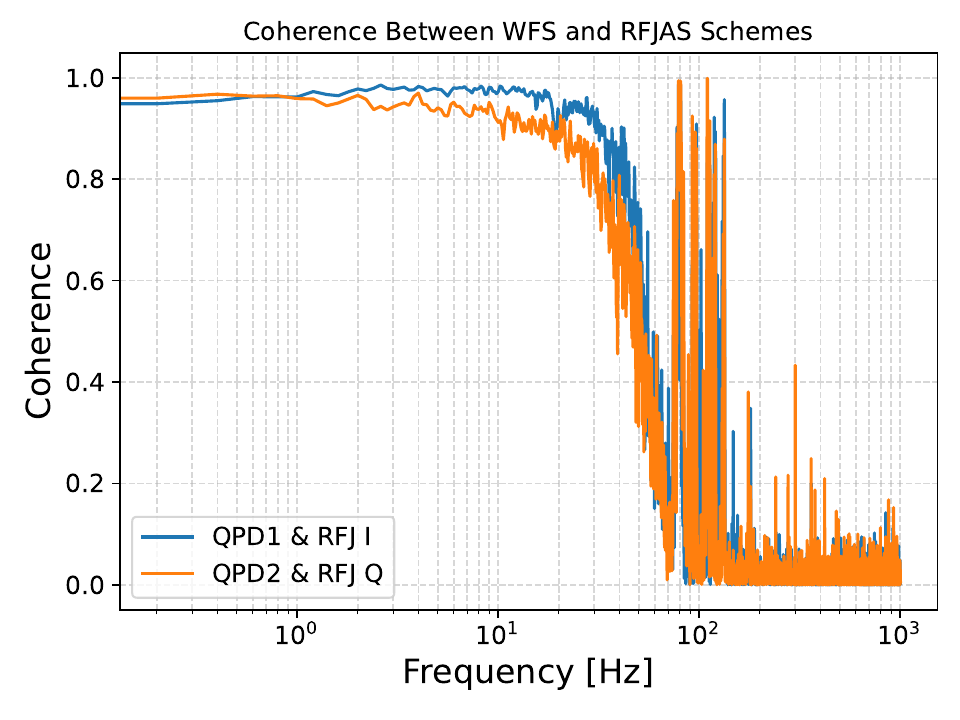}
    \caption{Measured coherence between the two sensing schemes. A high coherence is observed up to around $30$\,Hz, after which it decreases.}
    \label{fig:coherence}
\end{figure}

We also plot the coherence between WFS QPD1 and the RFJAS I-channel, and between WFS QPD2 and the RFJAS Q-channel. The coherence between the two sensing schemes remains close to 1 at frequencies up to approximately 30\,Hz in both alignment DOFs, after which it steeply declines. The results are shown in Fig.~\ref{fig:coherence}.

This behavior is expected, given that both sensing schemes have a relatively high signal-to-noise ratio up until 30\,Hz, at which point the WFS scheme becomes limited by electronics noise.

\section{Conclusion} \label{sec:conclusion}

In this work, we studied two alignment sensing schemes that differ in how they detect the alignment signal.
Wavefront sensing (WFS) utilizes two quadrant photodetectors (QPDs) that are separated by \(90^\circ\) of Gouy phase, while RFJAS utilizes a single element photodetector with two orthogonal RF demodulation channels, namely I and Q Channels.
WFS is sensitive to beam spot motion on the QPDs, which couples into alignment signals differently at different mode mismatches, while RF jitter alignment sensing (RFJAS) is largely insensitive to beam position. However, RFJAS can suffer from RF amplitude modulation (RFAM), which appears as offsets in the alignment signal.

Our tabletop experiments show that the currently used WFS scheme is limited by beam spot motion (BSM) at frequencies below $\sim30$\,Hz, which coincides with the frequency range where LIGO is limited by alignment noise \cite{Capote_2025}. Capote \cite{Capote_thesis} has measured the ASC noise budget at the LIGO Hanford detector, but the specific contribution of BSM to this budget has not yet been quantified. On site, LIGO employs DC centering loops to maintain beam centering on the QPDs, which is an important difference from our tabletop setup. However, these loops have their own limitations and contribute noise themselves, so a sensing scheme that does not rely on them could be advantageous. While the RFJAS scheme could in principle be affected by BSM through clipping and finite aperture effects, this can be mitigated by ensuring the beam is well contained within the active area of the RFPD. 

We therefore propose that a similar analysis to ours be carried out for the WFS scheme at the LIGO site to determine the contribution of BSM to the ASC noise budget. If BSM is indeed a limiting factor for the interferometer, then RFJAS may be a better option for aLIGO ASC. It is possible however that the shot-noise limited sensitivity for RFJAS is worse than for WFS. Under the assumption that both schemes should be shot noise limited at higher frequencies, a blended alignment sensing scheme that operates RFJAS at low frequencies and WFS at higher frequencies could provide the best solution. Implementing RFJAS would require two beam deflectors modulating independently the alignment in the tangential and sagittal planes, adding some hardware complexity, but it remains technically feasible.

Future work includes testing a blended scheme in practice, as well as closing alignment loops for both alignment schemes. A dedicated theoretical study into the expected noise performance of both schemes in the aLIGO interferometers is also a critical next step in this work.
Overall, the combined use of RFJAS and WFS offers a promising path to improved alignment sensing in gravitational-wave detectors such as aLIGO.

\section{Acknowledgments}
This work was supported by NSF awards PHY-2110360 and PHY-2409530.

\appendix
\section{Shot Noise Calculation} \label{appendix:shot_noise}
To compute the shot noise for the noise budget, we measure the DC power incident on the RFPD and the QPDs and use Eqn.~\ref{eq:shot_nosie}. 
\subsection*{RFPD}
For RFPD, the incident power is 3.5\,mW. The RFPD is made of InGaAs which has a high quantum efficiency, therefore we do not take that into account in the calculation.
Plugging the power back into Eqn.~\ref{eq:shot_nosie}:

\begin{equation}
    \sqrt{S_{\text{sn}}(f)} =
    \sqrt{2 \times \hbar \omega \times 3.5 \times 10^{-3}} 
    \approx 2.55 \times 10^{-11} \,\mathrm{W}/\sqrt{\mathrm{Hz}}
\end{equation}

The RFPD has a optical gain of 3\,V/mW. Using the diagonalized matrices combinations, we can express the shot noise in \(\frac{|HG_{10}/HG_{00}|}{\sqrt{Hz}}\), thus it is directly comparable to the noise budget terms.

\begin{align}
\sqrt{S_{\text{I}}(f)} &\approx
  7.65\times10^{-8}\,\frac{V_s}{\sqrt{\mathrm{Hz}}}
  \times 0.0079\,\frac{|HG_{10}/HG_{00}|}{V_d} \nonumber \\
  &\quad \times \tfrac{1}{0.0042}\,\frac{V_d}{V_s} \nonumber \\
  &= 1.43\times10^{-7}\,\frac{|HG_{10}/HG_{00}|}{\sqrt{\mathrm{Hz}}}, \\[6pt]
\sqrt{S_{\text{Q}}(f)} &\approx
  7.65\times10^{-8}\,\frac{V_s}{\sqrt{\mathrm{Hz}}}
  \times 0.01\,\frac{|HG_{10}/HG_{00}|}{V_d} \nonumber \\
  &\quad \times \tfrac{1}{0.00332}\,\frac{V_d}{V_s} \nonumber \\
  &= 2.30\times10^{-7}\,\frac{|HG_{10}/HG_{00}|}{\sqrt{\mathrm{Hz}}}.
\end{align}
Therefore, the total shot noise contribution 

\begin{equation*}
\text{RFJAS}_{\text{SN}} = \sqrt{(1.43 \times 10^{-7})^2 +(2.30 \times 10^{-7})^2 }
\end{equation*}

\vspace{-0.5em} 

\begin{equation}
\text{RFJAS}_{\text{SN}}= 2.71 \times 10^{-7} \,\frac{|HG_{10}/HG_{00}|}{\sqrt{\mathrm{Hz}}}
\end{equation}

\subsection*{QPDs}
Since the QPDs consist of left and right quadrants, and it is incoherent, the total shot noise in the WFS signal is the quadrature sum of the shot noise of each quadrant. 

\begin{align*}
    P_{tot} &= P_L + P_R \\
    P_{WFS} &= P_L - P_R
\end{align*}
Now, the shot noise of WFS is 

\begin{align*}
    \tilde{P}_{WFS} = \sqrt{\tilde{P_L^2}+\tilde{P_R^2}} \\
    \tilde{P}_{WFS} = \sqrt{2 \times \hbar \omega  P_L+2 \times \hbar \omega  P_R}
\end{align*}
which then becomes

\begin{align}
    \tilde{P}_{WFS}=\sqrt{2 \times \hbar \omega  P_{tot}}
    \label{equ:shot_noise_WFS}
\end{align}
identical to that of a single element photodetector. Moreover, the QPDs are made out of silicon, which has a quantum efficiency of about \(20\%\) at 1064\,nm wavelength. 
Therefore, with 1.75 mW incident on each QPD:

\begin{equation}
    \sqrt{S_{\text{sn}}(f)} \times \sqrt{1/0.2}\approx 4.02 \times 10^{-11} \,\mathrm{W}/\sqrt{\mathrm{Hz}}
\end{equation}
Similar to the RFJAS case, the QPDs have an optical gain of about \(1.8 \frac{V}{mW}\). Therefore,

\begin{align}
\sqrt{S_{\text{QPD1}}(f)} &\approx
  7.24\times10^{-8}\,\frac{V_s}{\sqrt{\mathrm{Hz}}}
  \times 0.0079\,\frac{|HG_{10}/HG_{00}|}{V_d} \nonumber \\
  &\quad \times \frac{1}{0.035}\,\frac{V_d}{V_s} \nonumber \\
  &= 1.62\times10^{-7}\,\frac{|HG_{10}/HG_{00}|}{\sqrt{\mathrm{Hz}}}, \\[6pt]
\sqrt{S_{\text{QPD2}}(f)} &\approx
  7.23\times10^{-8}\,\frac{V_s}{\sqrt{\mathrm{Hz}}}
  \times 0.01\,\frac{|HG_{10}/HG_{00}|}{V_d} \nonumber \\
  &\quad \times \frac{1}{0.0455}\,\frac{V_d}{V_s} \nonumber \\
  &= 1.62\times10^{-7}\,\frac{|HG_{10}/HG_{00}|}{\sqrt{\mathrm{Hz}}}.
\end{align}

Finally, the total contribution to shot noise of both QPDs is the quadrature sum 

\begin{equation}
    S_{WFS}=\sqrt2 \times 1.6\times 10^{-7} \approx 2.26\times 10^{-7} \,\frac{|HG_{10}/HG_{00}|}{\sqrt{\mathrm{Hz}}}
\end{equation}
Which is very close to that of a single element photodetector, as expected from Equ.~\ref{equ:shot_noise_WFS}

From all above, and as seen in the noise budget in Fig.~\ref{fig:Noise budget}, we see that the shot noise is not limiting our measurements. Since it is below all of the other noises, we are not plotting it and we suffice with this calculation.


\bibliography{apssamp}

\end{document}